\documentclass[twocolumn,showpacs,preprintnumbers,amsmath,amssymb]{revtex4}

\usepackage{graphicx}
\usepackage{dcolumn}
\usepackage{bm}

\begin{document}

\preprint{APS/123-QED}

\title{A Rich Example of Geometrically Induced Nonlinearity: From Rotobreathers
and Kinks to Moving Localized Modes and Resonant Energy Transfer}

\author{P. G. Kevrekidis $^1$, S. V. Dmitriev $^{2,3}$, S. Takeno$^4$,
A. R. Bishop$^5$ and E. C. Aifantis$^6$}
\affiliation{ $^1$Department of Mathematics and Statistics,
University of Massachusetts Lederle Graduate Research Tower,
Amherst, MA 01003-4515, USA \\
$^2$Institute of Industrial Science, the University of Tokyo, 4-6-1 Komaba, Meguro-ku, Tokyo 153-8505, Japan \\
$^3$National Institute of Materials Science, 1-2-1 Sengen, Tsukuba, Ibaraki 305-0047, Japan \\
$^4$Graduate School, Nagasaki Institute of Applied Science, Nagasaki 851-0193, Japan \\
$^5$Center for Nonlinear Studies and Theoretical Division, Los Alamos National Laboratory, Los Alamos, NM 87545 USA \\
$^6$Laboratory of Mechanics and Materials, Aristotle University of
Thessaloniki, GR 54124, Thessaloniki, Greece \\
}
\date{\today}

\begin{abstract}
We present an experimentally realizable, simple mechanical system
with linear interactions whose geometric nature leads to
nontrivial, nonlinear dynamical equations. The equations of motion
are derived and their ground state structures are analyzed.
Selective ``static'' features of the model are examined in the
context of nonlinear waves including rotobreathers and kink-like
solitary waves. We also explore ``dynamic'' features of the model
concerning the resonant transfer of energy and the role of moving
intrinsic localized modes in the process.
\end{abstract}

\pacs{05.45.-a, 05.45.Yv, 63.20.-e}

\maketitle

\section{ Introduction.}

In the past few years, there has been a dramatic
increase of interest in the behavior of solitary waves and intrinsic
localized modes (ILMs) in dynamical lattices; see e.g., \cite{review} for
a number of recent reviews. One of the key reasons for this
focus of interest has been the ability of such modes (which are ubiquitous
in nonlinear lattice models) to localize the energy and transfer it in a
targeted way \cite{target1,target2}. An additional, related
feature of these modes is the important role they play in the conduction
of heat (or equivalently transport of energy) along such dynamical
simple lattices and how this relates to fundamental macroscopic laws
of thermodynamics such as Fourier's law of heat conduction;
see e.g.,  \cite{fourier2} for a recent review. Another application
may be found in modelling the deformation and fracture behaviour
of continuous structured media with internal degrees of freedom.
Some initial model equations in this direction can be found in \cite{Aifantis}.

On the other hand, an increasingly important theme in nonlinear physics
concerns the interplay between nonlinear dynamics and geometry,
especially in lattice settings. The relevant contexts vary from long-range
interactions on a fixed curved substrate \cite{gaid}, to substrate-feedback
models \cite{pgk2} and coupled atomic chains \cite{konchain}, and from
junctions between lattices with different masses \cite{bena} to
semi-circular polymer-like chains \cite{tsir} and models of geometrically
nontrivial DNA \cite{ACMG01}. The unifying principle in all these
situations is that the geometry can significantly affect the static
properties of excitations in the lattice (e.g., multi-stability), as well as
dynamical ones (e.g., a variety of outcomes in the interaction of
intrinsic localized modes with curvature).

Motivated by these two emerging themes of nonlinear lattice
dynamical systems, we propose in this work a mechanical example,
in which even though the underlying interactions of the system
consist of {\it linear} springs, the geometry renders the
interactions nonlinear. This, in turn, leads to complex
features of the straightforwardly realizable (in a mechanical
experiment) system at hand. Such examples are manifested in the
coherent structures of the system ranging from the familiar
kinks to the more exotic rotobreathers \cite{roto} (or other structures
such as domain walls or kinks with embedded defects \cite{dmitriev}).
They are also evident in the resonant energy transport features
of the model that relate the energy conduction with the moving
intrinsic localized modes (ILMs) that are present in the model
(see below). We touch upon each of the above aspects to
give a flavor of the rich and diverse properties that such
a ``geometrically-induced nonlinearity'' model is endowed with.

\section{ Model }

\begin{figure}
\includegraphics{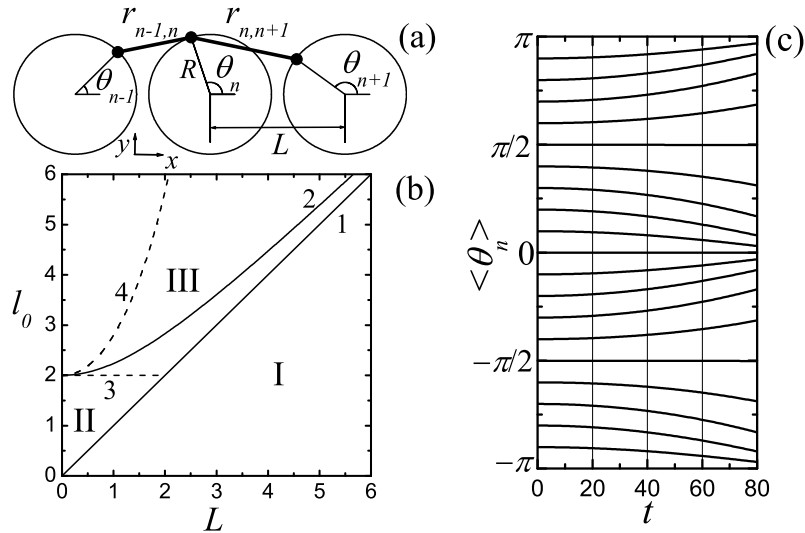}
\caption{(a) The bead-ring configuration: the rings of radii $R$
are at distance $L$ between them (either on the same or on
different planes). The dynamical variable of interest for each
particle is its azimuthal angle $\theta_n$ and the chain has a
nearest neighbor coupling through linear, elastic springs of
natural length $l_0$. The distance between adjacent particles is
denoted by $r_{n,n+1}$. (b) The parameter space of the model
$(L,l_0)$ divided into three regions with different ground state
structures. (c) Illustration of dynamical instability in the
regime I for all $\phi$ except for $\phi=0$ and $\phi=\pm \pi$.
Average atomic positions $<\theta_n>$ for the chain of $N=400$
particles are shown as the functions of time. Initially the
particles are placed at $\theta_n=\phi$ with different magnitudes
of $\phi$, and a small amplitude random perturbation is introduced
in the particle positions to initiate their vibrations.}
\label{dfig1}
\end{figure}

The mechanical example that we examine consists of masses (``beads'')
that slide on {\it fixed} rings of radius $R$ (even though the radius
can be variable from ring to ring, we will here consider it to be fixed).
Furthermore, the centers of the rings are at distance $L$ between them.
The chain of beads moving azimuthally along their respective rings
is coupled through linear, elastic strings of a natural length $l_0$;
see Fig. \ref{dfig1}(a) for a schematic.
There are two interesting subcases. The rings can be in the same plane
or they can be in different planes. While the latter case is also
of interest, we restrict ourselves to the former (coplanar rings) in what
follows. Then,
\begin{eqnarray}
x_n=n L + R \cos{\theta_n}, \,\,\,\,\,\,
y_n=R \sin(\theta_n),
\label{eqn2}
\end{eqnarray}
and the distance between adjacent particles is given by:
\begin{eqnarray}
r_{n,n+1} &=& \sqrt{(x_{n+1}-x_n)^2 + (y_{n+1}-y_n)^2}.
\label{Rnn1a}
\end{eqnarray}
The model Hamiltonian for linear, elastic interactions reads:
 \begin{eqnarray}
H=\sum_{n}\left[
\frac{1}{2}MR^2\left(\frac{\rm{d}\theta_n}{\rm{d}t}\right)^2
+\frac{1}{2}K\left(r_{n,n+1}-l_0\right)^2
\right].
\label{Hamiltonian}
\end{eqnarray}
Out of the 5 model parameters (particle mass $M$, spring constant $K$,
disk radius $R$, distance between disk centers $L$ and string natural length
$l_0$), two ($M, K$) can be scaled out while out of the remaining three
length scales, one can be used as a measure for the others (hence we
set $R=1$ in what follows). The resulting equations of motion read:
\begin{eqnarray}
\ddot{\theta}_n &=& \left[\sin(\theta_{n+1}-\theta_n)
- \sin(\theta_n-\theta_{n-1}) \right]
\nonumber
\\
&+& l_0 L \sin(\theta_n)
\left(\frac{1}{r_{n,n+1}} - \frac{1}{r_{n-1,n}} \right)
\nonumber
\\
&-& l_0 \left( \frac{\sin(\theta_{n+1}-\theta_n)}{r_{n,n+1}}
                 - \frac{\sin(\theta_{n}-\theta_{n-1})}{r_{n-1,n}} \right),
\label{EqOfMotion}
\end{eqnarray}
with $r^2_{n,n+1}=L^2 + 2
-2 \cos(\theta_{n+1}-\theta_n)
+ 2 L [\cos(\theta_{n+1}) - \cos(\theta_n)]$.
For convenience, we introduce the notation $L_0=\sqrt{L^2+4}$.

{\it Ground States.} The model of Eq. (\ref{EqOfMotion}) supports
numerous complex
structures (a number of which we will examine below),
but there are only three types of the ground state structures
bearing a very simple form.
The parameter space $(L,l_0)$ is naturally divided into three
regions, depending on the corresponding types of
ground state structure.
We now summarize these structures (see also Fig. \ref{dfig1}(b)).

For $l_0<L$, we are in regime I, where the lowest energy structure is
\begin{eqnarray}
\theta_n=\phi ,
\label{Type1}
\end{eqnarray}
for constant $\phi$.
Note that from static considerations, structures with any $\phi$
are in indifferent equilibrium, but in the presence of dynamic
perturbations, only structures with $\phi=0,\pi$ are stable.
The physical origin of the instability for $\phi \neq \{0,\pi\}$ can be easily
understood considering the situation with all particles being at rest
with $\theta_n = \phi$ and one particle oscillating with a finite
amplitude near the equilibrium position. Taking into account higher order
anharmonic terms, one can demonstrate an asymmetry
in the torques acting on the particle with respect to deviation
from the right versus deviation from the left. This asymmetry gives rise
to the force driving the particle toward the closest stable position, namely
$\phi=0$ or $\phi=\pi$, where the symmetry is restored. There are
two more symmetric structures corresponding to $\phi=\pi/2$ and
$\phi=-\pi/2$, but they are unstable. To illustrate this issue, for the chain
of $N=400$ particles, we set the initial conditions $\dot{\theta}_n=0$,
$\theta_n=\phi+r_n$, with different magnitudes of $\phi$ and $r_n$
being a random number homogeneously distributed on $[-0.05,0.05]$.
In Fig. \ref{dfig1}(c) we plot the time evolution of
$<\theta_n>=N^{-1}\sum_n \theta_n$.
One can see that there are two stable positions ($\phi=0$ and $\phi=\pi$)
with respect to dynamic fluctuations.

We thus restrict our considerations to the
stable equivalent structures, $\phi=0,\pi$.
In this case, the dispersion relation reads:
\begin{eqnarray}
\omega(k)=2\sqrt{1-\frac{l_0}{L}} \sin (\pi k) .
\label{Spectrum1}
\end{eqnarray}
In the limit $l_0=L$, the linear spectrum collapses to a single
point (in this case $\omega=0$). Such geometrically induced limits
where the linear spectrum collapses to a single point for all $k$
are an interesting feature of the present model (see below) and will
be called {\it purely anharmonic} (PA) limits, as the harmonic linear
part of the spectrum is eliminated in this case. This is a situation
that bears some resemblence to the anti-continuum limit of models of
ILMs \cite{review}.

Regime II is defined by $L<l_0<L_0$, where structures of the form:
\begin{eqnarray}
\theta_n=\pm [(-1)^n\phi +\delta \pi ],
\,\,\,\,\,\sin^2\phi=\frac{l_0^2-L^2}{4},
\label{Type2}
\end{eqnarray}
have zero energy (all springs have their
natural length); $\delta=0,\pm 1$. In this case, the linear spectrum is
of the form:
\begin{eqnarray}
\omega(k)=\sqrt{l_0^2-L^2} \sqrt{L^2+(4-l_0^2)\sin^2(\pi k)} ,
\label{Spectrum2}
\end{eqnarray}
which for $l_0=2$ becomes $k$-independent (PA limit),
$\omega(k)=L\sqrt{4-L^2}$.
The spectrum of Eq. (\ref{Spectrum2}) vanishes at
$l_0=L$ which is the border between the type I and type II regimes.

Finally, regime III consists of the natural lengths such that $L_0<l_0$,
when the ground state structure is
\begin{eqnarray}
\theta_n=\pm(-1)^n\frac{\pi}{2}.
\label{Type3}
\end{eqnarray}
In this case, the linear spectrum is given by
\begin{eqnarray}
\omega(k)=2 \sqrt{\left(1-\frac{4l_0}{L_0^3}\right)
\sin^2\left(\frac{\pi}{2}-\pi k\right)+\frac{l_0}{L_0}-1}.
\label{Spectrum3}
\end{eqnarray}
One can see that for $l_0=L_0^3/4$,
the width of linear spectrum vanishes (PA limit), $\omega(k)=L$.

Having discussed the ground states, we now give representative examples of the
interesting nonlinear behavior that is possible in each of the regimes highlighted above.

\section{ Regime I}
\label{RegimeI}

\begin{figure}
\includegraphics{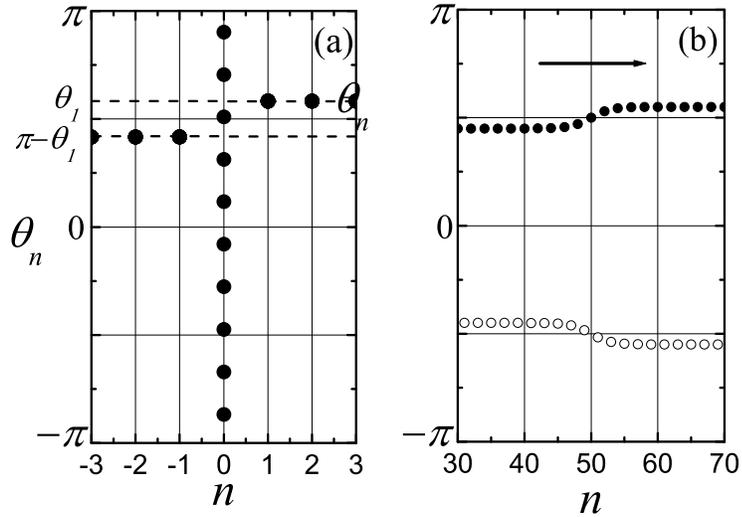}
\caption{(a) Numerical solution for a rotobreather in regime I.
The $n=0$ particle rotates with angular velocity $\omega=20$ at
$L=3$ and $l_0=1$. We show the stroboscopic picture of motion with
intervals $T/10$ (a tenth of the period $T$). Particles with
positive and negative $n$ are practically at rest at
$\theta_n=\pi/2\pm 0.261$. Numerically we found the root of Eq.
(\ref{Integral}) at $\theta_1=\pi/2+0.259$. (b) A kink solution is
shown in  regime II for $v=0.2$, $L=1$,
$l_0=\sqrt{4\sin^2\phi+L^2}$, with $\phi=(9/20)\pi$ (close to
$\phi=\pi/2$). Even and odd particles are shown by filled and open
 circles, respectively. Even in this case of (not very big) kink width,
Eq. (\ref{phi4kinkdiscrete}) provides a very good approximation.
The arrow shows the direction of propagation of the kink. }
\label{dfig2}
\end{figure}

Regime I supports an interesting rotobreather with one rotating
particle. Such an example is shown  in
Fig. \ref{dfig2}(a), where the particle
$n=0$ rotates with angular velocity $\omega=20$ at $L=3$ and $l_0=1$.
We show the stroboscopic picture of motion with intervals $T/10$
(where $T$ is the period of the solution).
Particles with positive and negative $n$ are practically at rest
at their respective positions.
The neighbors of $n=0$ particle can be at rest
in the positions where the averaged (over a period) torque acting from
the moving particle is equal to zero. For a large rotobreather frequency,
the angular velocity of the $n=0$ particle is almost constant and, hence,
the  averaging over time can be substituted by averaging over angle
$\theta_0$.
Thus, an approximate rotobreather solution can be given as follows:
\begin{eqnarray}
\theta_{n}=\omega t\,,\,\,\,\,\,\,\,\,\,\,{\rm for}\,\,\,\,n=0\, ,  \nonumber \\
\theta_{n}=\pm\theta_{1}\,,\,\,\,\,\,\,\,\,{\rm for}\,\,\,\,n>0\, ,\nonumber \\
\theta_{n}=\pm(\pi-\theta_{1})\,,\,\,\,\,{\rm for}\,\,\,\,n<0\, ,
\label{RotobreatherI}
\end{eqnarray}
i.e., the 0th particle moves with constant angular velocity $\omega$
while all other particles are at rest at
the positions defined through the angle $\theta_{1}$, which is a root of
\begin{eqnarray}
\int_{0}^{2\pi}M_{1}(\theta_{0} ,\theta_{1})d\theta_{0} = 0 ,
\label{Integral}
\end{eqnarray}
where $M_{1}(\theta_{0} ,\theta_{1})$ is the torque acting on the particle
$n=1$ from the
neighboring particles under the assumption that
$\theta_2=\theta_1$ and $\theta_0$
is arbitrary.
While the solutions of Eq. (\ref{Integral}) with $\theta_{1}=0,\pi$
lead to unstable configurations,
the root lying in the interval  $\pi/2<\theta_{1}<\pi$ leads to a
stable rotobreather which very accurately captures our numerical
findings (see the caption of Fig. \ref{dfig2}(a)).

Note that in the case of roto-breather presented in Fig. \ref{dfig2}(a),
dynamic fluctuations are absent. Introduction of such fluctuations would result
in the appearance of a weak forces driving the particles with $n<0$
to the position $\theta_n=0$ and particles with $n>0$ to the positions
with $\theta_n=\pi$ [see Fig. \ref{dfig1}(c)]. An equilibrium state in this case
would look similar to what is presented in Fig. \ref{dfig2}(a) only
in the vicinity of the roto-breather, but more distant particles
would approach the positions which are stable against dynamic
fluctuations.

\section{ Regime II }
\label{RegimeII}

As two prototypical examples of the structures that can be supported
in regime II, we derive a kink solution and provide a numerical
example of a highly localized, moving ILM.

{\it Kink solution.} Using
\begin{eqnarray}
\theta_n=(-1)^n\left(\frac{\pi}{2}+\varepsilon_n\right) ,
\label{ansatz1}
\end{eqnarray}
where $\varepsilon_n \ll \pi$ and assuming that
$\varepsilon_n$ varies slowly with $n$, we obtain from Eq.
(\ref{EqOfMotion}) [up to cubic terms]
\begin{eqnarray}
\ddot{\varepsilon}_{n} &=&
\frac{B}{L^2}(\varepsilon_{n-1}-2\varepsilon_{n}+\varepsilon_{n+1})
+C\varepsilon_n
\nonumber
\\
&-& \frac{D}{16}\left[(\varepsilon_{n+1}+\varepsilon_{n})^3+(\varepsilon_{n}+\varepsilon_{n-1})^3\right],\
\label{discrete}
\end{eqnarray}
with
$B=L^2\left(1-\frac{4l_0}{L_0^3}\right)$,
$C=4\left(1-\frac{l_0}{L_0}\right)$
and $D=\frac{8}{3}\left(1-\frac{l_0(L^2+1)}{L_0^3}\right)$.
In the continuum limit, Eq. (\ref{discrete})
reduces to the $\varphi^4$-equation,
\begin{eqnarray}
\varepsilon_{tt}=
B\varepsilon_{xx}+C\varepsilon-D\varepsilon^3 .
\label{phi4}
\end{eqnarray}
When $C>0$ and $D>0$, the background potential has a  double-well structure
and Eq. (\ref{phi4}) supports topological solitons (kinks
and antikinks) \cite{belova}.
When the kink width is much greater than $L$, the solution
of Eq. (\ref{phi4}) can be used to approximate
the kink solution of the discrete Eq. (\ref{EqOfMotion}):
\begin{eqnarray}
\theta_{n}=(-1)^n\left[
\frac{\pi}{2}\pm\sqrt{\frac{B}{D}}\tanh\left[Q(nL-vt)\right]
\right],
\label{phi4kinkdiscrete}
\end{eqnarray}
where $v<\sqrt{B}$ is the kink velocity and $Q=\sqrt{C/[2(v^2-B)]}$.
We have verified that even for relatively small kink widths (i.e., moderate
discreteness), Eq. (\ref{phi4kinkdiscrete}) approximates well the
numerically obtained kink solutions [see Fig. \ref{dfig2}(b)].

We note in passing that in regime II there exist
various different types of domains, as illustrated in
Eq. (\ref{Type2}). Hence, there are possibilities to create
additional kinks, connecting different steady states than
the ones presented above.

\begin{figure}
\includegraphics{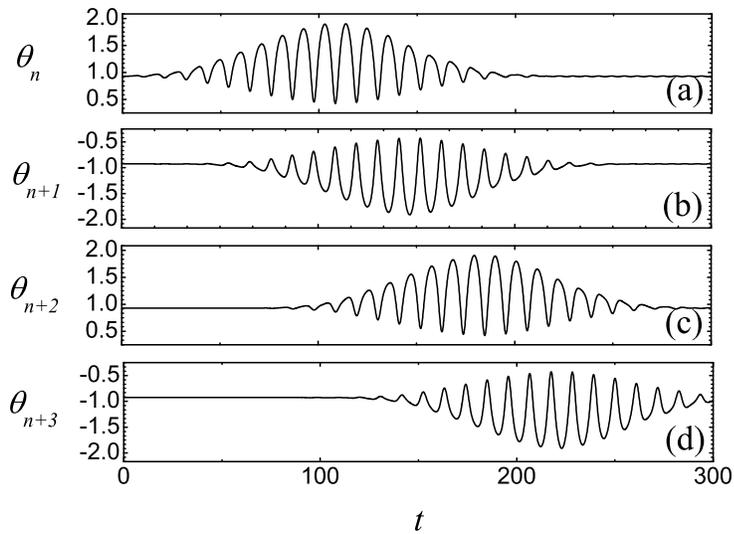}
\caption{Moving ILM in the regime II for $l_0=2$ (PA limit) and
$L=1.2$. } \label{dfig3}
\end{figure}

{\it Moving ILM.} An interesting example of a moving ILM is
presented in Fig. \ref{dfig3} for $l_0=2$ (PA limit), $L=1.2$. The
localized mode emerged from the local perturbation introduced to
$n=0$ particle of the structure $\theta_n=(-1)^n\phi$ by setting
$\theta_0=\phi+\pi /2$ with zero initial velocities for all
particles at $t=0$. The asymmetry in the displacements of
particles renders this structure less tractable analytically. We
will give a very simple analytical approximation for the ILMs
(standing and moving) in regime III where the displacements are
symmetric. However, we believe that the mechanism of ILM
propagation in the regime II is the same as in the regime III,
i.e., the relay-like resonant energy exchange between particles.

\section{ Regime III }
\label{RegimeIII}

{\it Standing and moving ILMs.}
Regime III supports ILMs which, depending on model parameters, can
be moving or standing. We have also found that the existence of standing ILMs
precludes the existence of moving ones and vice versa. To better
understand this
phenomenon we first solve an auxiliary problem of resonant energy transfer and
then give some simple analytical estimates for standing and moving ILMs.
The robustness of moving ILMs against their collisions is
also verified numerically.

Here we consider the PA limit, $l_0=L_0^3/4$.
If a particle is forced according to:
\begin{eqnarray}
\theta_0(t)=\frac{\pi}{2}+A\sin(\omega t),
\label{Drive}
\end{eqnarray}
with $A \ll \pi$, we are interested in the motion
of its nearest neighbors, $\theta_{-1}(t)=\theta_1(t)=-\frac{\pi}{2}-\varepsilon(t)$
with $\varepsilon \ll \pi$, assuming that all other particles
are at rest at their equilibrium positions. Retaining
up to linear terms in $\varepsilon$ and cubic in $A$,
we obtain
\begin{eqnarray}
\ddot{\varepsilon}+\omega_0^2\varepsilon=
\beta A^2\sin^2(\omega t)+\gamma A^3\sin^3(\omega t), \nonumber \\
\omega_0=L,\,\,\,
\beta=\frac{L(2-L^2)}{4L_0^2},\,\,\,
\gamma=\frac{2-L^2+2L^4}{3L_0^4}-\frac{1}{6},
\label{EqOfMotionSmal}
\end{eqnarray}
with the particular solution
\begin{eqnarray}
\varepsilon(t)=
\frac{\beta A^2}{2\omega_0^2}
+\frac{3\gamma A^3}{4(\omega_0^2-\omega^2)}\sin(\omega t) \nonumber \\
-\frac{\beta A^2}{2(\omega_0^2-4\omega^2)}\cos(2\omega t)
-\frac{\gamma A^3}{4(\omega_0^2-9\omega^2)}\sin(3\omega t),
\label{ParticularSolution}
\end{eqnarray}
which gives the first three resonance harmonics. When the forced
particle oscillates with a nearly resonant frequency, the
amplitude of its neighbors can grow significantly.

Equation (\ref{EqOfMotionSmal}) does not take anharmonicity into account.
The latter effect  was studied
numerically, where we found that the anharmonicity is hard,
i.e., the oscillation frequency of the $n=1$ particle grows
as a function of amplitude,  for $L<L^{\ast}\approx 1.65$,
and the situation is reversed for $L>L^{\ast}$.

The dynamics of the chain with one forced particle
differs qualitatively for hard and soft anharmonicity.
For $L<L^{\ast}$, when the free particle has
maximum amplitude, it oscillates in phase with the forced particle,
and its amplitude excceeds $A$ in the resonance regime
while for $L>L^{\ast}$ it does not because the particles
oscillate out of phase. In other words, efficient inter-particle
energy exchange occurs only for hard anharmonicities.
For the chain with soft anharmonicity, any local perturbation
remains local.

We carry out the following numerical experiments.
We excite a single site  according
to Eq. (\ref{Drive}) for times $0\le t\le \tau$ with $\tau=5000$
and calculate the power of the energy source, $W=E/\tau$, where $E$ is
the total energy of the chain at $t=\tau$. The chain is long enough so that
at $t=\tau$ the perturbation produced by the forced particle has not reached
the boundaries.

\begin{figure}
\includegraphics{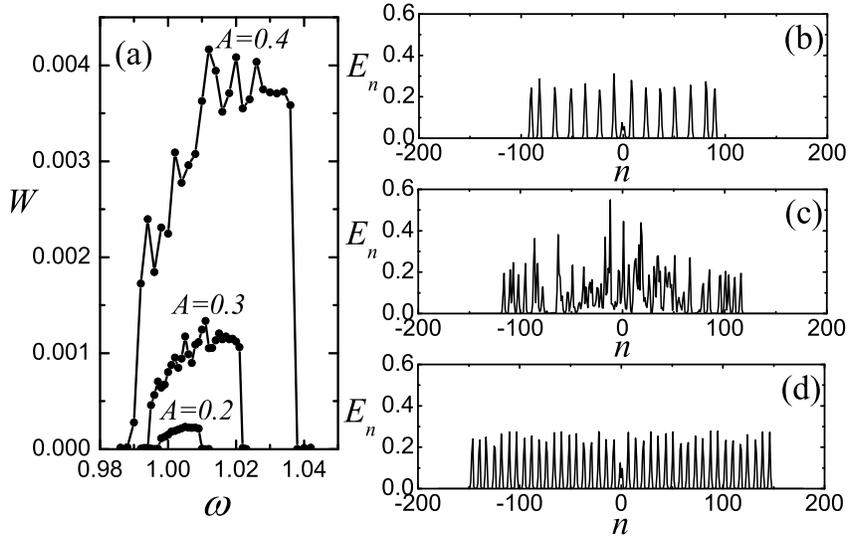}
\caption{(a) The power of the energy source $W$ is shown as a
function of the driving frequency $\omega$ for $L=1$ and for
$A=0.2,\,\,0.3,\,\,0.4$. Right Panel: the distribution of energy
at $t=\tau$. Model parameters are $L=1$, $l_0=L_0^3/4$, $A=0.4$
and (b) $\omega=0.992$, (c) $\omega=1.020$, (d) $\omega=1.033$.
The panel shows the particle energies $E_n$, averaged over the
period $2\pi/\omega$. In (b) and (d) the forced particle emits
ILMs periodically, while in (c) it emits chaotically. }
\label{dfig4}
\end{figure}

We have found that $W$ can be nonzero only for $L<L^{\ast}$,
regardless of the magnitudes of $A$ and $\omega$
in Eq. (\ref{Drive}). The mechanism of the energy transfer
is the emission of moving ILMs. In Fig. \ref{dfig4}(a),
we present $W$ as  functions of $\omega$
for $L=1$ and for $A=0.2,\,\,0.3,\,\,0.4$. It can be seen
that the smaller  $A$ is, the narrower the window of $W>0$.
The figure also shows the distribution of energy
in the chain at $t=\tau$ for $L=1$, $l_0=L_0^3/4$,
and driving parameters
$A=0.4$ and (b)
$\omega=0.992$, (c) $\omega=1.020$, (d) $\omega=1.033$.
In (b) and (d) (edges of the window
with $W>0$), the forced particle emits ILMs periodically
while in (c) (central part of the window) chaotically.
For $L>L^{\ast}$, there is no efficient energy exchange
between particles and moving ILMs are not possible.
Instead, stable standing ILMs arise that
are localised at the excited particle. An approximate
solution can be expressed by a conventional perturbation
method assuming that only one particle moves:
\begin{eqnarray}
\theta_0 (t)=\frac{\pi}{2}+A_1\sin(\omega t)+A_3\sin(3\omega t) , \nonumber \\
\omega^2=\omega_0^2+\frac{3}{4}BA_1^2 ,\,\,\,\,\,\,\,
A_3=\frac{-BA_1^3}{32\omega_0^2+27BA_1^2} ,
\label{StandingILM}
\end{eqnarray}
where $\omega_0^2=\frac{4l_0(L^2+2)}{L_0^3}-2$,
$B=\frac{1}{3}-l_0\frac{5L^6+8L^4+16L^2+16}{3L_0^7}$.

\begin{figure}
\includegraphics{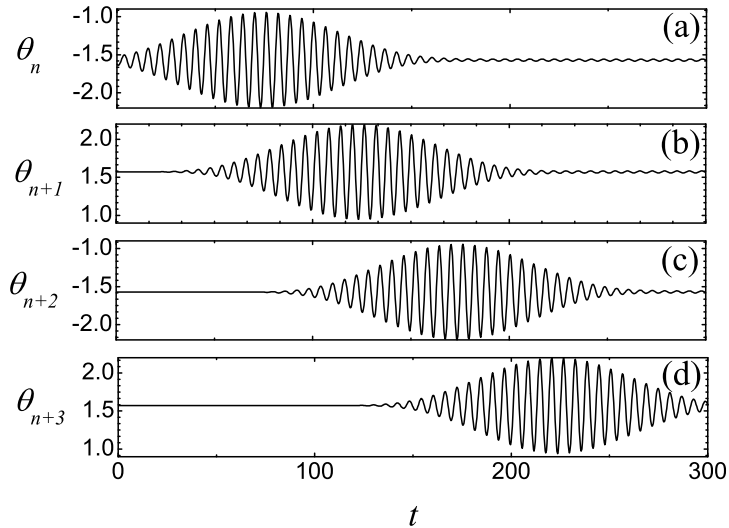}
\caption{A moving ILM is shown by the functions $\theta_n(t)$ for
the four nearest nodes. Particles show the relay-like motion
oscillating near the equilibrium positions
$\theta_n=(-1)^n(\pi/2)$. Model parameters are $l_0=L_0^3/4$ (PA
limit in regime III), and $L=1$ ($<L^\ast$). The parameters in the
solution Eq. (\ref{MovingILM}) are $A_1=0.6$, $a=0.4$. Marginal
radiation can be seen after the ILM passes a node (at large
times). The ILM propagates rather slowly, it travels one lattice
spacing $L$ in about $8T$. } \label{dfig5}
\end{figure}

The moving ILM is (practically) localized at three particles:
\begin{eqnarray}
\theta_{n-1}=aA_1\sin\left(\omega t\pm\frac{2\pi}{3}\right) , \nonumber \\
\theta_{n}=A_1\sin\left(\omega t\right)+A_3\sin\left(3\omega t\right) , \nonumber \\
\theta_{n+1}=aA_1\sin\left(\omega t\mp\frac{2\pi}{3}\right) ,
\label{MovingILM}
\end{eqnarray}
where upper and lower signs correspond to ILM moving in positive and
negative directions, respectively. The amplitude $A_1$ of the ILM
is a free parameter. The relation between $A_1$, the ILM frequency
$\omega$, and the third harmonic amplitude, $A_3$, is given
by Eq. (\ref{StandingILM}). Empirically we have found that
$a=0.4$ gives a good result over a wide range of ILM amplitudes
(see Fig. \ref{dfig5}).

\begin{figure}
\includegraphics{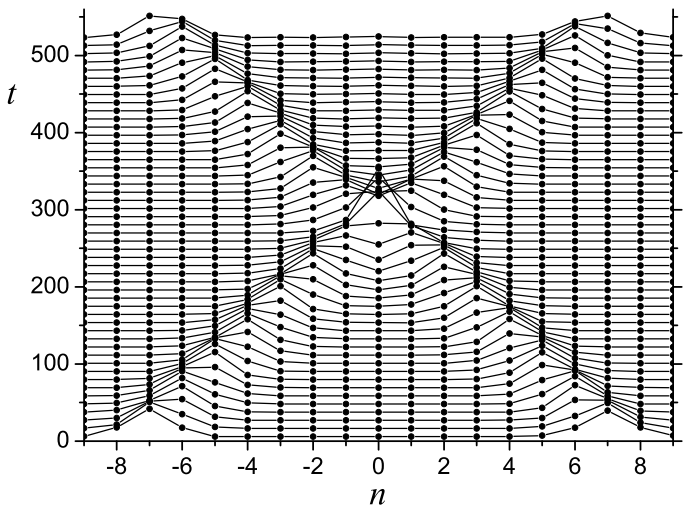}
\caption{Elasticity of the in-phase collision of two moving ILMs
defined by Eq. (\ref{MovingILM}) with $A=0.7$, $a=0.4$. The model
parameters are $L=1$ ($<L^\ast$) and $l_0=L_0^3/4$ (PA limit). }
\label{dfig6}
\end{figure}

In Fig. \ref{dfig6} we show an in-phase collision of two ILM
defined by Eq. (\ref{MovingILM}) with $A=0.7$, $a=0.4$. Model
parameters are $L=1$, $l_0=L_0^3/4$ (PA limit). Shown are the
snapshots of $H_n$ at different time $t$, where $H_n$ is the total
energy of $n$th particle (kinetic and potential). We can see that
such moving ILMs can interact with each other in a quasi-elastic
fashion.

\begin{figure}
\includegraphics{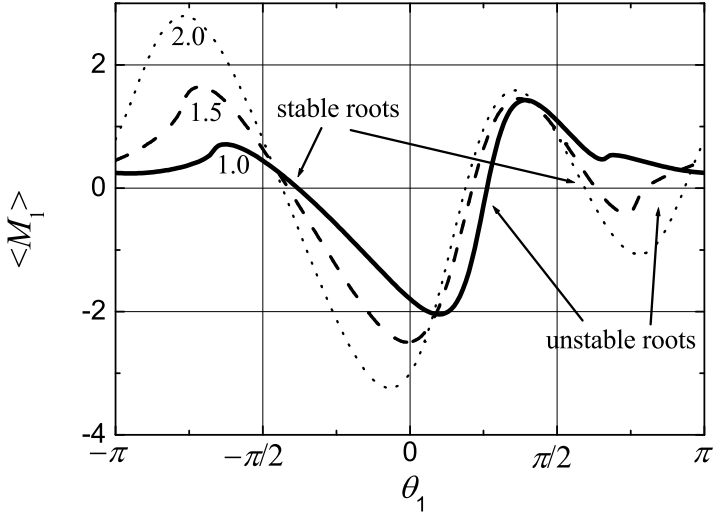}
\caption{Torque (averaged over the angle $\theta_0$) acting from
rotating particle $(n=0)$ on its nearest neighbor $(n=1)$ which is
at rest at $\theta_1$ for $l_0=L_0^3/4$ (PA limit) and $L=2$
(dotted), $L=1.5$ (dashed), $L=1$ (solid). } \label{dfig7}
\end{figure}

\begin{figure}
\includegraphics{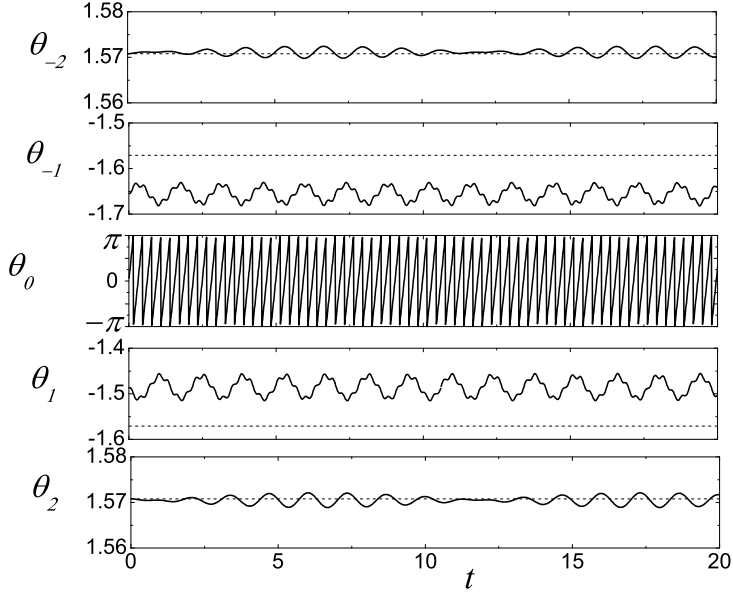}
\caption{Rotobreather in the regime III initiated by setting
initial angular velocity $\dot\theta_0=20$ and initial positions
of the $n=\pm 1$ particles $\theta_\pm 1=-\pi/2 \pm 0.084$
corresponding to zero averaged torque acting from $n=0$ particle.
Model parameters are $L=5$ and $l_0=L_0^3/4$ (PA limit). Dashed
horizontal lines show $\theta_n=\pm\pi/2$. } \label{dfig8}
\end{figure}

\begin{figure}
\includegraphics{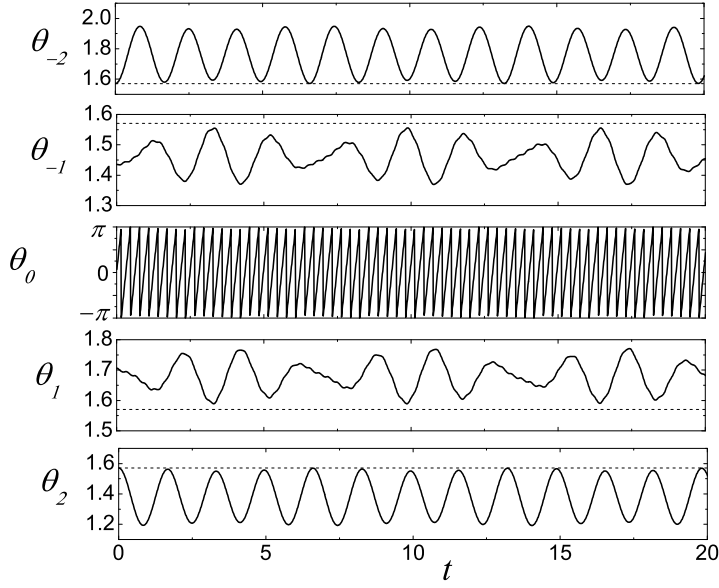}
\caption{Rotobreather corresponding to the root of equation
$<M_1>=0$ which appeares at large $L$. In this case $L=5$,
$l_0=L_0^3/4$ (PA limit) and the root was found for $n=\pm 1$
particles at $\pi/2\pm 0.135$. Note that here the particles $n=\pm
1$ oscillate near $\pi/2$ but not near $-\pi/2$ as in the case
presented in Fig. \ref{dfig8}. This is because the stable root
which appeares at $L>1.3$ is shifted by, roughly, $\pi$ compared
to the root existing for all $L$ (see Fig. \ref{dfig7}). }
\label{dfig9}
\end{figure}

{\it Rotobreathers.} Regime III also supports a rotobreather with
one rotating particle. Here again, as in Sec. \ref{RegimeI}, we
assume that for a rotobreather with sufficiently large frequency
the torque acting from the rotating particle on its nearest
neighbors can be estimated by averaging over angle. In Fig.
\ref{dfig7} we show the (averaged over angle) torque acting from
rotating ($n=0$) particle on its nearest neighbor ($n=1$), which
is assumed to be at rest at $\theta_1$, for $l_0=L_0^3/4$ (PA
limit) and $L=2$ (dotted), $L=1.5$ (dashed), $L=1$ (solid). We are
interested in the positions $\theta_1$ corresponding to zero
torque. For $L<1.3$ there are only two roots (one of them is
stable) and for $L>1.3$ there are four roots (two of them are
stable). For $L=5$, for example, the stable root was found at
$\theta_1=-\pi/2+ 0.084$ which is in a good agreement with what is
observed for the rotobreather presented in Fig. \ref{dfig8}. For
$L=5$, another stable root for the $n=1$ particle was found at
$\theta_\pm 1=\pi/2+0.135$ and the existance of this rotobreather
was also confirmed numerically (see Fig. \ref{dfig9}).

\section { Conclusions}

In this paper we have presented a novel nonlinear dynamical system,
consisting of an easily realizable mechanical example where the
nonlinearity is induced by the geometry of the problem. We have illustrated
the laws of motion and the rich static, dynamic (both equilibrium and
non-equilibrium) behavior of the system. We have identified some of the
relevant coherent structures including kink-like heteroclinic connections
and roto-breathing periodic orbits and have seen some of the interesting
dynamical phenomenology including the ``conducting'' (for hard anharmonicities)
or  ``insulating'' (for soft anharmonicities) behavior of the system and
the role of moving or standing ILMs, respectively, as energy carriers.
It would be of interest to examine further from an analytical (as well as
from a numerical or experimental) perspective the phenomenological wealth
of such a model. Such studies are currently in progress and will be reported
in future publications.

\end{document}